# Nitrogen doping of carbon nanoelectrodes for enhanced control of DNA translocation dynamics

Sang Won Jung,[1] Han Seul Kim,[2] Art E. Cho,[1,*] Yong-Hoon Kim[2,*]

[1] Department of Bioinformatics, Korea University, Sejong, 2511 Sejong-ro Jochiwon-eup, Sejong 30019, Korea.

[2] Graduate School of Energy, Environment, Water, and Sustainability, Korean Advanced Institute of Science and Technology, 291 Deahak-ro, Yuseong-gu, Daejeon 34141, Korea.

Controlling the dynamics of DNA translocation is a central issue in the emerging nanopore-based DNA sequencing. To address the potential of heteroatom doping of carbon nanostructures to achieve this goal, herein we carry out atomistic molecular dynamics simulations for single-stranded DNAs translocating between two pristine or doped carbon nanotube (CNT) electrodes. Specifically, we consider the substitutional nitrogen doping of capped CNT (capCNT) electrodes and perform two types of molecular dynamics simulations for the entrapped and translocating single-stranded DNAs. We find that the substitutional nitrogen doping of capCNTs stabilizes the edge-on nucleobase configurations rather than the original face-on ones and slows down the DNA translocation speed by establishing hydrogen bonds between the N dopant atoms and nucleobases. Due to the enhanced interactions between DNAs and N-doped capCNTs, the duration time of nucleobases within the nanogap was extended by up to ~ 290 % and the fluctuation of the nucleobases was reduced by up to ~ 70 %. Given the possibility to be combined with extrinsic light or gate voltage modulation methods, the current work demonstrates that the substitutional nitrogen doping is a promising direction for the control of DNA translocation dynamics through a nanopore or nanogap based of carbon nanomaterials.

## 1. Introduction

Toward the goal of accomplishing an effective next-generation DNA sequencing scheme that will realize personalized or precision medicine,[1-4] solid-state nanopores have emerged as a promising platform and significant research effort has been devoted into their development in the past few years.[3-12] Compared with the more established bio-nanopores, solid state nanopores can allow the enhanced device stability and application of advanced fabrication technology available in semiconductor microelectronics industry. Their advantages also include the possibility to utilize readout methods other than longitudinal ionic currents and/or to more effectively incorporate novel low-dimensional nanomaterials such as graphene and carbon nanotubes (CNTs). For example, due to the atomically thin structure and unique electrical transport properties of graphene[13] and CNTs,[14] graphene nanopores[15-18] and nanogaps[19-22] can possibly allow the detection of DNA sequence based on in-plane transverse electrical currents and achieve the coveted single-molecule resolution.

To realize the full potential of solid-state DNA sequencing based on novel low-dimensional nanomaterials and be able to read nucleobases at the high spatial resolution, in addition to the improved readout scheme and electronics, a high-fidelity control of DNA translocation dynamics still needs to be achieved.[5, 7, 9, 23, 24] Specifically, one of the main challenges is to reduce the speed of DNA translocation through a nanopore or nanogap. In the current technologies, the electrophoretic translocation speed of single-stranded DNA (ssDNA) is still too high (typically around 10 base/μs, being $10^3 - 10^4$ times higher than that of biological pores) to correctly differentiate nucleobase at the single-base resolution.[6] Moreover, if the molecular conformation within a nanopore/nanogap can be precisely controlled such that optimal nucleobase configurations for readout are induced and structural fluctuations are minimized, it should achieve the greatly increased sensing accuracy by allowing the enhanced signal and minimized noise, respectively. While significant experimental advances have been made for the control of DNA translocation speed,[25-31] they were mostly limited to the conventional nanopores made of inorganic materials such as SiN and the advance in the nanopores/nanogaps based on graphene and related two-dimensional materials have been rather slow.[32, 33] To make rapid progress in this field, as in the case for the conventional nanopore counterparts,[24, 34-38] atomistic molecular dynamics (MD) simulations are expected to become a valuable aid by providing various microscopic information and device design guidelines.[20, 33, 39-44] For example, it was suggested that the reduction of the ssDNA translocation speed can be achieved by the hydrogenation of graphene edges,[20] electrical biasing,[41] or forming multilayer graphene layers.[42] In this regard, a promising yet so far unexplored direction is the heteroatom doping, which is an important modification process of the sp² -bonded carbon network that has been proven to be beneficial for various potential device applications,[45-49] and significant advances have been already made for their experimental realization.[50]

In this work, performing atomistic simulations, we study how the introduction of substitutional nitrogen dopant atoms into the sp² carbon network affects the DNA translocation dynamics. As the carbon-based nano-constriction platform (Figures 1a and 1b), we employ CNT caps for which we have previously shown that the nitrogen doping significantly enhances the magnitude of tunneling currents as well as the chemical selectivity of different nucleobases.[51] Regarding the choice of CNT cap geometry, we emphasize that the nanoscale curvature makes CNT and fullerenes distinctively different from other graphitic carbons in their ability to adsorb molecules.[52] In this context, due to the resemblance between CNT caps and fullerenes, the results obtained here can be directly applied to the ssDNAs translocating on the fullerene-functionalized graphene,[53] CNT sidewall,[54] and CNT tip[55] ("nanobuds"). In view of the carbon curvature effect, another DNA sequencing platform where our study is relevant is the nanopores or nanoelectrodes based on graphitic polyhedral or folded graphene edges,[56] for which the slowing down of DNA translocation speed has been already demonstrated experimentally.[32] In terms of the device architecture, we envision that, as schematically shown in Figure 1c, such one-dimensional nanogap can be embedded into a two-dimensional nanopore platform by, e.g., forming a buckypaper.[57]

Carrying out density functional theory (DFT) calculations for the capCNT and N-doped capCNT (N-capCNT), we will first show that significant and nontrivial localized electronic modifications are induced from the substitutional N doping. Incorporating the DFT data, we then carry out force-fields (FF) equilibrium and non-equilibrium atomistic MD simulations for the ssDNAs entrapped and translocating between (N-)capCNTs, respectively. The comparisons between capCNT- and N-capCNT-based MD simulations demonstrate that the substitutional nitrogen doping facilitates the instant formation of hydrogen bonds between nucleobases and the N dopant atoms within N-capCNT electrodes, and results in the desired slowdown of DNA translocation speed and better-defined edge-on nucleobase configurations (aromatic rings of nucleobases are aligned along the CNT axial direction with their functional groups heading toward the CNT caps; see Figure 1b) with reduced DNA fluctuations. Our work thus demonstrates that the heteroatom doping of carbon nanoelectrodes is a promising direction in view of controlling DNA conformations as well as reading nucleobases during ssDNA translocation processes.[8, 51]

## 2. Computational Methods

### 2.1 Models and FF MD simulations

FF MD simulations were performed using the NAMD program within the periodic boundary condition.[58] CHARMM27 force field parameters were used for DNA,[59, 60] TIP3P water molecules,[61] and ions.[62] The parameters of carbon atoms for benzene in CHARMM27 force field were used for those of carbon atoms in capCNT.[60, 63] The parameters for the carbon atoms bonded with the nitrogen atom in N-capCNT were those of type CG2R64, and the corresponding nitrogen atoms were those of type NG2R62, which were designed for 1,3,5-triazine.[45] Mulliken partial charges calculated with DFT were assigned to (N-)capCNT. Sodium and chloride ions were added into the system as a concentration of 0.15 M. The final system measured $96 \times 48 \times 120$ Å³ in size and included about 54,000 atoms. The MD integration time step used was 1 fs, and the particle-mesh Ewald (PME) scheme[64] with grid density of 1/Å³ was employed for the descriptions of electrostatics. Van der Waals energies were calculated using a 12 Å cutoff. A Langevin thermostat was assumed to maintain constant temperature at 295 K. The CNTs were fixed during all the simulations. The two sets of simulations were first energy minimized for 10,000 steps, followed by gradual heating from 0 to 295 K in 5 ps and solvent equilibration for 5 ps. The production run for the *Set 1* simulations ran up to 300 ps and the *Set 2* simulations up to 30 ps. For the *Set 2* simulations, 420 mV/Å of external electric field, which is comparable to that employed in experiments,[65] was applied to drive the translocation of ssDNA. Visualization and structure analysis were performed using the VMD software packages.[66]

### 2.2 DFT Calculations

To extract Mulliken charge distributions within (N-)capCNT electrodes, we carried out DFT calculations within the Perdew-Burke-Ernzerhof parameterization of generalized gradient approximation[67] using the SIESTA software.[68] Troullier-Martins-type norm-conserving pseudopotentials are employed to remove the core electrons.[69] The Kohn-Sham orbitals were expanded in terms of the double-ζ-plus-polarization level numerical atomic basis sets constructed with the confinement energy of 80 meV. The exchange-correlational energy and Hartree potential were calculated with a mesh cut-off energy of 300 Ry. CNT caps were introduced at the end of six-unit-long (5,5) CNT. A sufficiently large vacuum region of 15 Å was included to avoid artificial interactions with neighboring images, and only the gamma point was sampled for Brillouin zone integration.

## 3. Results and Discussion

### 3.1 Electron distribution within pristine and N-doped CNT caps

As the nano-constriction platform, we employed capped (5,5) CNTs with the diameter of 6.8 Å, which is shorter than the spacing of ~ 7 Å between adjacent nucleobases in a stretched ssDNA (Figure 2a).[51] N-capCNTs were prepared by substituting a carbon atom among the 5-membered-ring at the tip of capCNT with an N dopant atom (Figure 2b). To address the effects of a substitutional N atom on the spatial charge distributions within the CNT cap, we compared the Mulliken charge populations of N-capCNT with those of capCNT. Recent experimental studies have shown that the nature of N-doped graphene can be rather complicated due to the diverse atomic bonding possibilities,[70] and we recently demonstrated that the Mulliken charge visualization is a powerful tool to develop the atomistic understanding of the bonding nature of N atoms embedded in an $sp^2$ carbon network.[71] A key finding from these studies was that there arise significant local charge fluctuations near the substitutional N atoms before a rather uniform net electron doping is achieved in the long-range region (beyond ~ 7 Å radius). For the N-doped graphene, we found the Mulliken population of $-0.402$ $e$ per nitrogen atom,[71] which is in good agreement with the experimentally estimated value of $-0.42$ $e$.[70]

We will now show that overall similar N-induced charge transfer characters are developed even within the curved geometry of CNT caps. First, considering the pristine CNT cap case, we observe that the presence of the curvature and five-membered rings already induce a finite spatial charge distributions but their magnitudes are very small (Figure 2a): While each atom at the first five-membered carbon ring equally donates electrons by $-0.001$ $e$, other atoms within the cap accept electrons by around $+0.004 \sim 0.005$ $e$ each. The net induced charge within the cap region was $+0.115$ $e$, and this is compensated by the tube region connected to the cap. On the other hand, in the N-capCNT case, we find that the magnitudes of charge distributions between the substitutional N and surrounding C atoms are several orders of magnitude larger (Figure 2b): The net Mulliken charge of the substitutional N dopant is $-0.349$ $e$, and most of these charges are transferred to the carbon atoms bonded to N, which results in the excess charge of $+0.071$ $e$ for C1 and C2 located within the pentagon ring and the larger amount of $+0.105$ $e$ for C3. The net charge within the cap region was $-0.065$ $e$ for the N-capCNT and again this is compensated by that of the tube region connected to the cap. We emphasize again that such strong electron donating nature of the substitutional N dopant is consistent with that found in the substitutional N-doped graphene.[70,71] Having observed that the introduction of a substitutional N atom within graphene or CNT caps induces a strongly localized charges, one could expect that they can experience drastically different and possibly enhanced electrostatic interactions with ssDNA compared with their undoped counterparts.

### 3.2 Conformational dynamics of ssDNA between N-doped CNT caps

We now move on to the FF MD simulation results, and discuss how the substitutional N doping CNT cap affects the conformation of nearby ssDNAs and particularly their translocation dynamics. To prepare junction models, two capCNTs or N-capCNTs were mirror-symmetrically placed with a gap distance of 12.0 Å which is large enough for the nucleobases to pass through in the edge-on configurations. Moreover, to systematically address the problem, we performed two sets of FF MD simulations using a ssDNA model composed of $A_1G_2C_3T_4A_5G_6C_7T_8$ (where A, G, C, T respectively stands for adenine, guanine, cytosine, and thymine) with the central four nucleobases $C_3T_4A_5G_6$ utilized for analysis: (1) First, we carried out equilibrium MD simulations for the ssDNA placed at the midpoint of nanogap and observed the impact of the substitutional N dopant atoms on the conformations and fluctuations of ssDNA (*Set 1*; see Supplementary Information Figure S1a and Movies S1-S4). (2) Next, by applying an electric field of 420 mV/Å along the translocation $z$ direction, we performed non-equilibrium steering MD simulations and studied the actual DNA translocation dynamics (*Set 2*; see Supplementary Information Figure S1b and Movies S5-S8). To ensure the reliability of the results, we repeated the calculations multiple times for each simulation set.

We first discuss the results from the equilibrium *Set 1* simulations. Here, to describe the orientations of DNA bases, a set of basis vectors was defined as follows (Figure 3, left panels): [40] We first defined a unit vector $e_1$ pointing from C4 to N1 atoms in purines (adenine and guanine) and from C6 to N3 in pyrimidines (cytosine and thymine). To determine $e_2$, we then defined a vector $e_2^*$ pointing from C4 to C6 in purines and from C6 to C4 in pyrimidines. The vector $e_2$ was then set as the unit vector proportional to $e_2^* - (e_2^* \cdot e_1) e_1$. The unit vector $e_3$ was finally defined as $e_3 \equiv e_1 \times e_2$. To describe these orientations more succinctly, we will use two angles $\alpha$ and $\beta$ (Figure 1b): The angle $\alpha$ was defined as the angle between the vectors $e_2^*$ and $y$ axis, in which $y$ is parallel to the CNT axis. The angle $\beta$ was defined as the angle between the vectors $e_3$ and $z$, where $z$ is normal to the CNT axis. In short, the $\beta$ angle quantitatively defines the preference of edge-on conformations of nucleobases, where $\beta = 0°$ and $\beta = 90°$ correspond to the ideal edge-on and face-on modes, respectively. On the other hand, the $\alpha$ angle defines the conformational preference of the long axis of nucleobases to align along the CNT axis.

In Figure 3 (right panels, see also Supplementary Information Figure S2), we show the *Set 1* MD simulations results. In the capCNT-based junctions, the average $\beta$ angles of deoxyadenosine 5′-monophosphate (dAMP), deoxyguanosine 5′- monophosphate (dGMP), deoxycytidine 5′-monophosphate (dCMP), and deoxythymidine 5′-monophosphate (dTMP) were 49 ± 35°, 63 ± 10°, 41 ± 13°, and 44 ± 12°, respectively. Namely, the nucleobases are found to preferably adopt face-on conformations with the tendency ordering of dGMP > dAMP > dTMP > dCMP. This trend can be understood in terms of the driving force to maximize π-π interactions between CNT cap and aromatic ring of nucleobases.[72] In line with the standard understanding on its nature[73, 74] and as will become more evident in the *Set 2* steering MD simulation results, these π-π interactions have partially repulsive characters. Overall, it is natural for purines with 9-membered double rings (dAMP and dGMP) interact more strongly to graphitic systems than pyrimidines with 5-membered rings (dTMP and dCMP). Indeed, for the nucleobases adsorbed on carbon-based systems such as $C_{60}$, single-wall carbon nanotubes (SWNTs), graphene, and graphite surface, the face-on tendency ordering of dGMP > dAMP > dTMP > dCMP was typically observed or calculated.[75-78]

On the other hand, we found that N-capCNTs induce significantly reduced $\beta$ angles or enhanced edge-on conformations for all four nucleobases: The average $\beta$ angles of dAMP, dGMP, dCMP, and dTMP were largely reduced from the capCNT cases to 24 ± 6°, 17 ± 8°, 28 ± 9°, and 18 ± 10°, respectively. Correspondingly, the preference ordering of dGMP > dTMP > dAMP > dCMP for the edge-on conformation (or dCMP > dAMP > dTMP > dGMP for the face-on conformation) is now different from that in the pristine capCNT case of dCMP > dTMP > dAMP > dGMP. These changes indicate that the π-π interactions, which were dominant for the pristine capCNT case, are now interrupted by hydrogen boding-induced attractions between the substitutional N atoms within CNT caps and the functional groups in nucleobases. The $\alpha$ angles (105 ± 4°, 60 ± 4°, 96 ± 8°, and 132 ± 7° for dAMP, dGMP, dCMP, and dTMP, respectively) provide detailed information on the source of hydrogen bonds for edge-on nucleobases. Specifically, we observed that the hydrogen bonds involving -C2-H in dAMP, -C2-N2-$H_2$ in dGMP, -C4-N4-$H_2$ and -C5-H in dCMP, and -C6-H and -C5-$CH_3$ in dTMP are established (see Supplementary Information Figure S3). Particularly, the maximization of edge-on conformations in the dGMP case can be understood by the fact that two functional groups, -N1-H and -C2-N2-$H_2$ on one side of guanine and -C8-H on the opposite side establish hydrogen bonds with the substitutional N dopant atoms on the two N-capCNTs in a linear geometry. Of course, its large molecular size should also help to establish stronger attractive interactions with N-capCNTs.

### 3.3 Electric field-driven translocation of ssDNA through N-doped CNT caps

Having analyzed the *Set 1* simulations that showed the clear advantage of substitutional N doping of capCNTs in controlling the nucleobase conformations (by effectively reducing the fluctuations of $\beta$ angle by more than 32 % and giving significant preference for edge-on modes due to the enhanced hydrogen bonding), we now discuss the *Set 2* steering MD simulations that will allow us to observe the realistic translocation dynamics in the existence of electrophoretic electric field (Figure 4; see also Supplementary Information Figure S4 and Movies S5-S8). Here, initial molecular geometries were prepared by displacing the $C_3$, $T_4$, $A_5$, and $G_6$ conformations between the N-capCNT electrodes obtained in *Set 1* simulations by +5.0 Å along the $z$ direction and additionally aligning them to assume nearly edge-on configurations ($\beta \sim 0°$).

Overall, we observed in the *Set 2* simulations the stepwise or ratcheting translocation, which was often observed in the graphene nanopore platform. [33, 41, 43, 44] Carrying out more detailed analysis, we first observe in the capCNT-based junctions that, irrespective of the initial geometries, the external electric field exerted along the translocation $z$ direction resulted in the dominantly face-on configurations ($\beta \sim 90°$) within the nanogap (marked by red shaded area). In addition to the π-π interactions discussed regarding *Set 1* simulations, this tendency should have been further strengthened by the additional interactions between the electrophoretic external electric field and net dipole moment of nucleobases (experimentally measured as 2.5, 7.1, 7.0, 4.1 Debye for A, G, C, and T, respectively).[5] In terms of the latter coupling between the nucleobase dipole moments and the electric field along the translocation direction, we point out that the face-on conformations should be common to all the nanopore sequencing platforms that involve electrophoretic DNA translocations. Here, while all the nucleobases eventually assume the face-on configurations before entering into the nanogap, we could observe that the size of nucleobases affect the speed of conformational change: Whereas the pyrimidine-based dCMP and dTMP were are abruptly rotated from the edge-on to face-on modes within only 5 ps, purine-based dAMP and dGMP were gradually aligned along the $z$ direction for about 10 ps period. Once they entered into the nanogap, we observe that the duration time of each nucleobase is also affected by the size of nucleobases and the smaller pyrimidines stay longer within the gap (retention time of ~ 6.5 ps) than their purine counterparts (retention time of ~ 4 ps). We interpret that the above differences in the nanogap entrance and exit behavior result from the partially repulsive characters of the π-π interactions.[73, 74]

Focusing now on the N-capCNT electrode case, we observe that it provides several advantages in terms of the translocation dynamics. Most notably, we observe that the nucleobases now enter the nanogap much faster than in the capCNT cases, reflecting the attractive forces exerted by N-doped CNT caps. Specifically, all the nucleobases entered the N-capCNT nanogap within 3 ps while the entry time was longer than 5 ps in the capCNT counterparts. Next, we observe that the additional attraction between nucleobases and N-doped CNT caps increased the dwell time within the nanogap. For example, the retention time of dAMP, which was 4 ps between capCNTs, now significantly increased up to more than 11 ps between N-capCNTs. In terms of nucleobase conformations, as can be expected from the *Set 1* simulations, they are now much closer to the edge-on mode with the $β$ angle of ~ 30 ° throughout all nucleobases (the $α$ angles are almost close to the *Set 1* simulation results). Note that here we did not include the effects of electric field along the CNT direction resulting from the bias voltages that can be applied to read the nucleobases based on transverse electrical currents, and we expect that its inclusion will further enhance the edge-on configurations and concurrently increase the retention time.

## 4. Conclusions

In summary, carrying out combined DFT-FF multiscale simulations, we demonstrated that the substitutional N doping of $sp^2$ carbon network significantly improves the controllability of DNA translocation dynamics, a critical element for the next-generation DNA sequencing. First, to prepare the FF simulation protocols, we carried out DFT calculations to identify the charge redistribution within CNT cap induced by the nitrogen doping. Then, we carried out equilibrium FF MD simulations and showed that the N doping of capCNTs not only significantly reduces the fluctuations of nucleobases but induces their predominant edge-on configurations via the additional hydrogen bonding and electrostatic interactions. Performing non-equilibrium steering MD simulations, we finally studied the actual electrophoretic ssDNA translocations through the nanogap between capCNT nanoelectrodes. For the case of pristine capCNT electrodes, we found that the translocation speed and conformations of ssDNA are dominated by the partially repulsive π-π interactions and external electric field, negatively affecting the nucleobase retention time with the nanogap. However, upon the N doping of capCNT electrodes, we found that the additional hydrogen bonding and electrostatic attractions notably enhances the controllability of ssDNA translocation by slowing down the DNA translocation speed and inducing dominantly edge-on conformations of nucleobases. Specifically, we found that the N doping of capCNTs reduces the translocation speed and fluctuations of ssDNA by up to 290 % and 70 %, respectively. Our work thus demonstrates that the substitutional heteroatom doping of carbon nanoelectrode provides an attractive route toward the enhanced controllability of DNA translocation dynamics. In closing, we emphasize that the *intrinsic* modification scheme proposed here can be straightforwardly combined with the *extrinsic* methods such as light[28] and/or gate voltage[27, 31] modulation to achieve better controllability and increase nanopore/nanogap dwelling time of the translocating DNA.

## Acknowledgements

This work has been supported by the Nano-Material Technology Development Program (Nos. 2016M3A7B4024133, 2016M3A7B4025405, and 2016M3A7B4909944) of the National Research

Foundation (NRF) funded by the Ministry of Science and ICT of Korea. H.S.K. and Y.-H.K. were additionally supported by the NRF Basic Research Program (No. 2017R1A2B3009872), Global Frontier Program (No. 2013M3A6B1078881), and Basic Research Lab Program (No. 2016M3A7B4909944). S.W.J. and A.E.C. were additionally supported by NRF Grant No. 2017R1D1A1B03035870.

## Author Information

### Corresponding Authors
*A.E.C.: artcho@korea.ac.kr
*Y.-H.K.: y.h.kim@kaist.ac.kr

## References

1. Branton, D.; Deamer, D. W.; Marziali, A.; Bayley, H.; Benner, S. A.; Butler, T.; Di Ventra, M.; Garaj, S.; Hibbs, A.; Huang, X.; Jovanovich, S. B.; Krstic, P. S.; Lindsay, S.; Ling, X. S.; Mastrangelo, C. H.; Meller, A.; Oliver, J. S.; Pershin, Y. V.; Ramsey, J. M.; Riehn, R.; Soni, G. V.; Tabard-Cossa, V.; Wanunu, M.; Wiggin, M.; Schloss, J. A. *Nat. Biotechnol.* **2008,** 26, 1146-1153.
2. Gupta, P. K. *Trends Biotechnol.* **2008,** 26, 602-611.
3. Metzker, M. L. *Nat. Rev. Genet.* **2010,** 11, 31-46.
4. Niedringhaus, T. P.; Milanova, D.; Kerby, M. B.; Snyder, M. P.; Barron, A. E. *Anal. Chem.* **2011,** 83, 4327-41.
5. Zwolak, M.; Di Ventra, M. *Rev. Mod. Phys.* **2008,** 80, 141-165.
6. Venkatesan, B. M.; Bashir, R. *Nat. Nanotechnol.* **2011,** 6, 615-24.
7. Yokota, K.; Tsutsui, M.; Taniguchi, M. *RSC Adv.* **2014,** 4, 15886-15899.
8. Kim, H. S.; Kim, Y. H. *Biosens. Bioelectron.* **2015,** 69, 186-198.
9. Li, J.; Yu, D.; Zhao, Q. *Microchimica Acta* **2015,** 183, 941-953.
10. Heerema, S. J.; Dekker, C. *Nat Nanotechnol* **2016,** 11, 127-136.
11. Lindsay, S. *Nat. Nanotechnol.* **2016,** 11, 109-111.
12. Di Ventra, M.; Taniguchi, M. *Nat Nanotechnol* **2016,** 11, 117-126.
13. Das Sarma, S.; Adam, S.; Hwang, E. H.; Rossi, E. *Rev. Mod. Phys.* **2011,** 83, 407-470.
14. Kang, Y.-J.; Kim, Y.-H.; Chang, K. J. *Curr. Appl. Phys.* **2009,** 9, S7-S11.
15. Nelson, T.; Zhang, B.; Prezhdo, O. V. *Nano Lett.* **2010,** 10, 3237-3242.
16. Saha, K. K.; Drndić, M.; Nikolić, B. K. *Nano Lett.* **2011,** 12, 50-55.
17. Girdhar, A.; Sathe, C.; Schulten, K.; Leburton, J.-P. *Proc. Natl. Acad. Sci. U.S.A.* **2013,** 110, 16748-16753.
18. Avdoshenko, S. M.; Nozaki, D.; Gomes da Rocha, C.; González, J. W.; Lee, M. H.; Gutierrez, R.; Cuniberti, G. *Nano Lett.* **2013,** 13, 1969-1976.
19. Postma, H. W. C. *Nano Lett.* **2010,** 10, 420-425.
20. He, Y.; Scheicher, R. H.; Grigoriev, A.; Ahuja, R.; Long, S.; Huo, Z.; Liu, M. *Adv. Funct. Mater.* **2011,** 21, 2674-2679.
21. Prasongkit, J.; Grigoriev, A.; Pathak, B.; Ahuja, R.; Scheicher, R. H. *Nano Lett.* **2011,** 11, 1941-1945.
22. Jeong, H.; Kim, H. S.; Lee, S.-H.; Lee, D.; Kim, Y.-H.; Huh, N. *Appl. Phys. Lett.* **2013,** 103, 023701.
23. Fyta, M.; Melchionna, S.; Succi, S. *J. Polym. Sci., Part B: Polym. Phys.* **2011,** 49, 985-1011.
24. Luan, B.; Martyna, G.; Stolovitzky, G. *Biophys. J* **2011,** 101, (9), 2214-22.
25. Fologea, D.; Uplinger, J.; Thomas, B.; McNabb, D. S.; Li, J. *Nano Lett.* **2005,** 5, 1734-1737.
26. de Zoysa, R. S.; Jayawardhana, D. A.; Zhao, Q.; Wang, D.; Armstrong, D. W.; Guan, X. *J. Phys. Chem. B* **2009,** 113, 13332-13336.
27. Tsutsui, M.; He, Y.; Furuhashi, M.; Rahong, S.; Taniguchi, M.; Kawai, T. *Sci. Rep.* **2012,** 2.
28. Di Fiori, N.; Squires, A.; Bar, D.; Gilboa, T.; Moustakas, T. D.; Meller, A. *Nat Nanotechnol* **2013,** 8, 946-951.
29. Larkin, J.; Henley, R.; Bell, D. C.; Cohen-Karni, T.; Rosenstein, J. K.; Wanunu, M. *ACS Nano* **2013,** 7, 10121-10128.
30. Krishnakumar, P.; Gyarfas, B.; Song, W.; Sen, S.; Zhang, P.; Krstic, P.; Lindsay, S. *ACS Nano* **2013,** 7, 10319-10326.
31. Liu, Y.; Yobas, L. *ACS Nano* **2016,** 10, 3985-3994.
32. Freedman, K. J.; Ahn, C. W.; Kim, M. J. *ACS Nano* **2013,** 7, 5008-5016.
33. Banerjee, S.; Wilson, J.; Shim, J.; Shankla, M.; Corbin, E. A.; Aksimentiev, A.; Bashir, R. *Adv. Funct. Mater.* **2015,** 25, 936-946.
34. Aksimentiev, A.; Heng, J. B.; Timp, G.; Schulten, K. *Biophys J* **2004,** 87, 2086-2097.
35. Lagerqvist, J.; Zwolak, M.; Di Ventra, M. *Nano Lett.* **2006,** 6, 779-782.
36. Lagerqvist, J.; Zwolak, M.; Di Ventra, M. *Biophys. J.* **2007,** 93, 2384-2390.
37. Luan, B.; Peng, H.; Polonsky, S.; Rossnagel, S.; Stolovitzky, G.; Martyna, G. *Phys. Rev. Lett.* **2010,** 104, 238103.
38. Kowalczyk, S. W.; Wells, D. B.; Aksimentiev, A.; Dekker, C. *Nano Lett.* **2012,** 12, 1038-44.
39. Sathe, C.; Zou, X.; Leburton, J.-P.; Schulten, K. *ACS Nano* **2011,** 5, 8842-8851.
40. Wells, D. B.; Belkin, M.; Comer, J.; Aksimentiev, A. *Nano Lett.* **2012,** 12, 4117-4123.
41. Shankla, M.; Aksimentiev, A. *Nat. Commun.* **2014,** 5, 5171-5171.
42. Liang, L.; Zhang, Z.; Shen, J.; Zhe, K.; Wang, Q.; Wu, T.; Ågren, H.; Tu, Y. *RSC Adv.* **2014,** 4, 50494-50502.
43. Zhang, Z.; Shen, J.; Wang, H.; Wang, Q.; Zhang, J.; Liang, L.; Ågren, H.; Tu, Y. *J. Phys. Chem. Lett.* **2014,** 5, 1602-1607.
44. Qiu, H.; Sarathy, A.; Leburton, J. P.; Schulten, K. *Nano Lett.* **2015,** 15, 8322-8330.
45. Yang, Y.; Li, X.; Jiang, J.; Du, H.; Zhao, L.; Zhao, Y. *ACS Nano* **2010,** 4, 5755-5762.
46. Dai, L. *Acc. Chem. Res.* **2013,** 46, 31-42.
47. Agnoli, S.; Granozzi, G. *Surf. Sci.* **2013,** 609, 1-5.
48. Maiti, U. N.; Lee, W. J.; Lee, J. M.; Oh, Y.; Kim, J. Y.; Kim, J. E.; Shim, J.; Han, T. H.; Kim, S. O. *Adv Mater.* **2014,** 26, 40-66.
49. Zhang, J.; Xia, Z.; Dai, L. *Sci Adv.* **2015,** 1, e1500564.
50. Wang, H.; Maiyalagan, T.; Wang, X. *ACS Catal.* **2012,** 2, 781-794.
51. Kim, H. S.; Lee, S. J.; Kim, Y.-H. *Small* **2014,** 10, 774-781.
52. Gotovac, S.; Honda, H.; Hattori, Y.; Takahashi, K.; Kanoh, H.; Kaneko, K. *Nano Lett.* **2007,** 7, 583-587.
53. Wu, X.; Zeng, X. C. *Nano Lett* **2009,** 9, 250-256.
54. Nasibulin, A. G.; Pikhitsa, P. V.; Jiang, H.; Brown, D. P.; Krasheninnikov, A. V.; Anisimov, A. S.; Queipo, P.; Moisala, A.; Gonzalez, D.; Lientschnig, G.; Hassanien, A.; Shandakov, S. D.; Lolli, G.; Resasco, D. E.; Choi, M.;

Tomanek, D.; Kauppinen, E. I. *Nat. Nanotechnol.* **2007,** 2, 156-161.
55. Choi, J. I.; Kim, H. S.; Kim, H. S.; Lee, G. I.; Kang, J. K.; Kim, Y. H. *Nanoscale* **2016,** 8, 2343-2349.
56. Tan, P.; Dimovski, S.; Gogotsi, Y. *Philos. Trans. A Math. Phys. Eng. Sci.* **2004,** 362, 2289-2310.
57. Wang, D.; Song, P.; Liu, C.; Wu, W.; Fan, S. *Nanotechnology* **2008,** 19, 075609.
58. Phillips, J. C.; Braun, R.; Wang, W.; Gumbart, J.; Tajkhorshid, E.; Villa, E.; Chipot, C.; Skeel, R. D.; Kale, L.; Schulten, K. *J. Comput. Chem.* **2005,** 26, 1781-802.
59. Foloppe, N.; MacKerell, J. A. D. *J. Comput. Chem.* **2000,** 21, 86-104.
60. MacKerell, A. D.; Banavali, N. K. *J. Comput. Chem.* **2000,** 21, 105-120.
61. Jorgensen, W. L.; Chandrasekhar, J.; Madura, J. D.; Impey, R. W.; Klein, M. L. *J. Chem. Phys.* **1983,** 79, 926-935.
62. Beglov, D.; Roux, B. *J. Chem. Phys.* **1994,** 100, 9050-9063.
63. He, H.; Scheicher, R. H.; Pandey, R.; Rocha, A. R.; Sanvito, S.; Grigoriev, A.; Ahuja, R.; Karna, S. P. *J. Phys. Chem. C* **2008,** 112, 3456-3459.
64. Darden, T.; York, D.; Pedersen, L. *J. Chem. Phys.* **1993,** 98, 10089-10092.
65. Merchant, C. A.; Healy, K.; Wanunu, M.; Ray, V.; Peterman, N.; Bartel, J.; Fischbein, M. D.; Venta, K.; Luo, Z.; Johnson, A. T. C.; Drndić, M. *Nano Lett.* **2010,** 10, 2915-2921.
66. Humphrey, W.; Dalke, A.; Schulten, K. *J. Mol. Graph.* **1996,** 14, 33-38.
67. Perdew, J. P.; Burke, K.; Ernzerhof, M. *Phys Rev Lett* **1996,** 77, 3865-3868.
68. Soler, J. M.; Artacho, E.; Gale, J. D.; García, A.; Junquera, J.; Ordejón, P.; Sánchez-Portal, D. *J. Phys. Condens. Matter* **2002,** 14, 2745-2779.
69. Troullier, N.; Martins, J. L. *Phys. Rev. B* **1991,** 43, 1993-2006.
70. Zhao, L.; He, R.; Rim, K. T.; Schiros, T.; Kim, K. S.; Zhou, H.; Gutierrez, C.; Chockalingam, S. P.; Arguello, C. J.; Palova, L.; Nordlund, D.; Hybertsen, M. S.; Reichman, D. R.; Heinz, T. F.; Kim, P.; Pinczuk, A.; Flynn, G. W.; Pasupathy, A. N. *Science* **2011,** 333, 999-1003.
71. Kim, H. S.; Kim, H. S.; Kim, S. S.; Kim, Y. H. *Nanoscale* **2014,** 6, 14911-14918.
72. Chen, X.; Rungger, I.; Pemmaraju, C. D.; Schwingenschlögl, U.; Sanvito, S. *Phys. Rev. B* **2012,** 85, 115436.
73. Hunter, C. A.; Sanders, J. K. M. *J. Am. Chem. Soc.* **1990,** 112, 5525-5534.
74. Curtis, M. D.; Cao, J.; Kampf, J. W. *J. Am. Chem. Soc.* **2004,** 126, 4318-4328.
75. Sowerby, S. J.; Cohn, C. A.; Heckl, W. M.; Holm, N. G. *Proc. Natl. Acad. Sci. U S A* **2001,** 98, 820-822.
76. Sun, W.; Bu, Y.; Wang, Y. *J. Phys. Chem. C* **2011,** 115, 3220-3228.
77. Wang, Y. *J. Phys. Chem. C* **2008,** 112, 14297-14305.
78. Antony, J.; Grimme, S. *Phys. Chem. Chem. Phys.* **2008,** 10, 2722-2729.

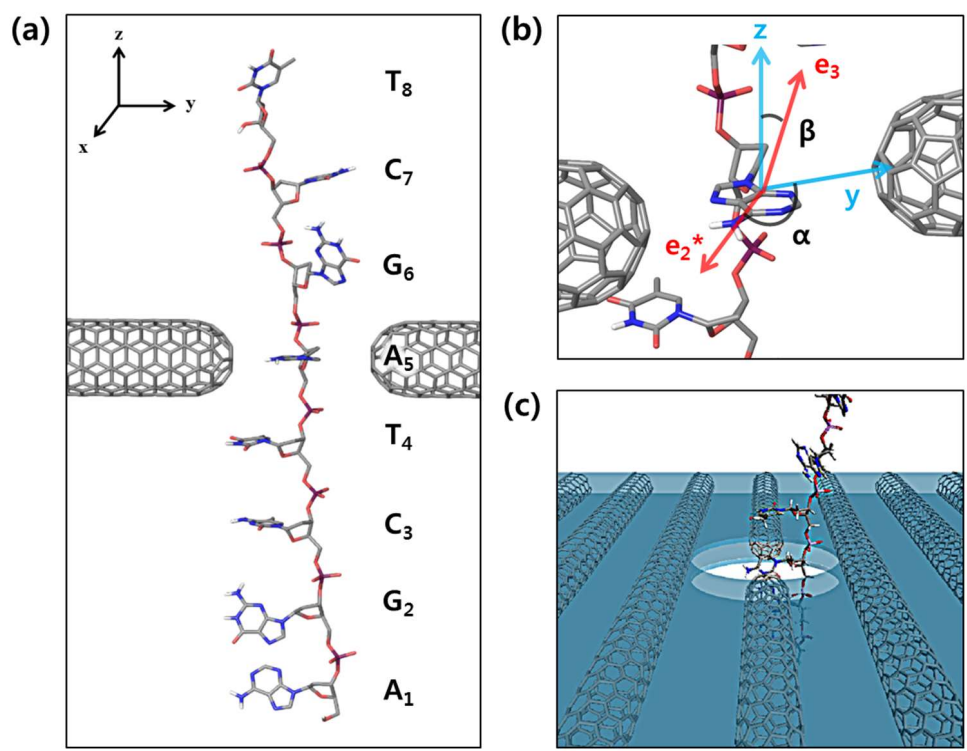

**Figure 1.** (a) A schematic of a CNT electrodes-based DNA sequencing device, where a stretched ssDNA segment ($A_1G_2C_3T_4A_5G_6C_7T_8$) is located between the electrodes. (b) Definition of the angles $\alpha$ and $\beta$ which are introducd to analyze the orientation of each nucleobase in *Set 1* and *2* simulations. $\alpha$ is defined as the angle between the vectors $e_2^*$ and y, where y is parallel to the cylindrical axis of (N-)capped CNTs. $\beta$ is defined as the angle between the vectors $e_3$ and z, where z is perpendicular to the cylindrical axis of (N-)capped CNT. (c) The illustration of buckypaper-based nanopore DNA sequencing device, which consists of CNT and its nanogap.

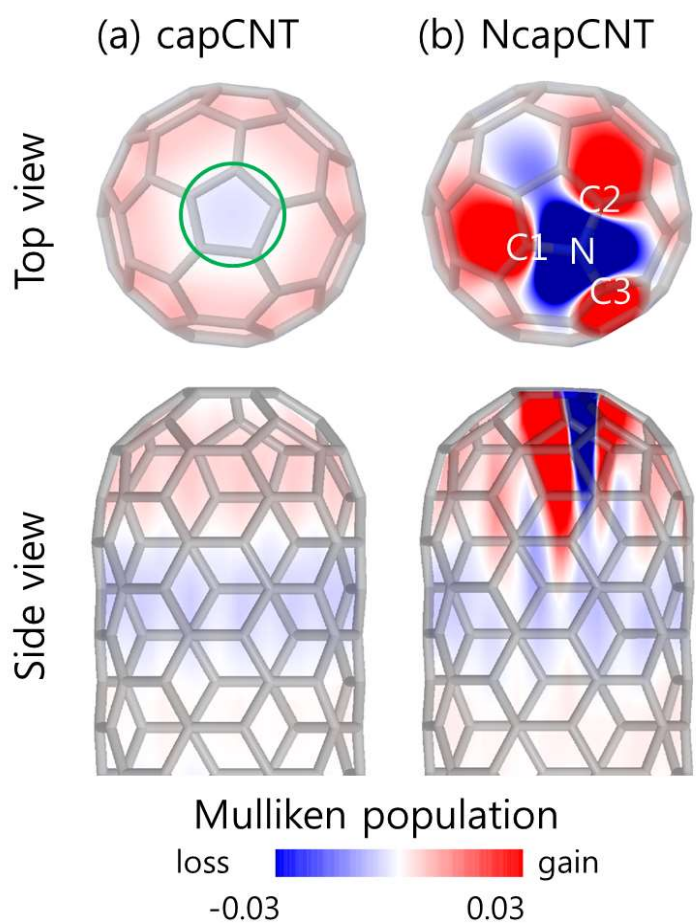

**Figure 2.** Charge distributions within the pristine and N-doped capped CNTs. Real-space projections of Mulliken populations are shown for the top (top panels) and side views (bottom panels) of (a) capCNT and (b) N-capCNT. Electron loss and gain are represented as blue and red colors, respectively. The green circle in the top panel of (a) indicates the first five-membered carbon ring located at the tip.

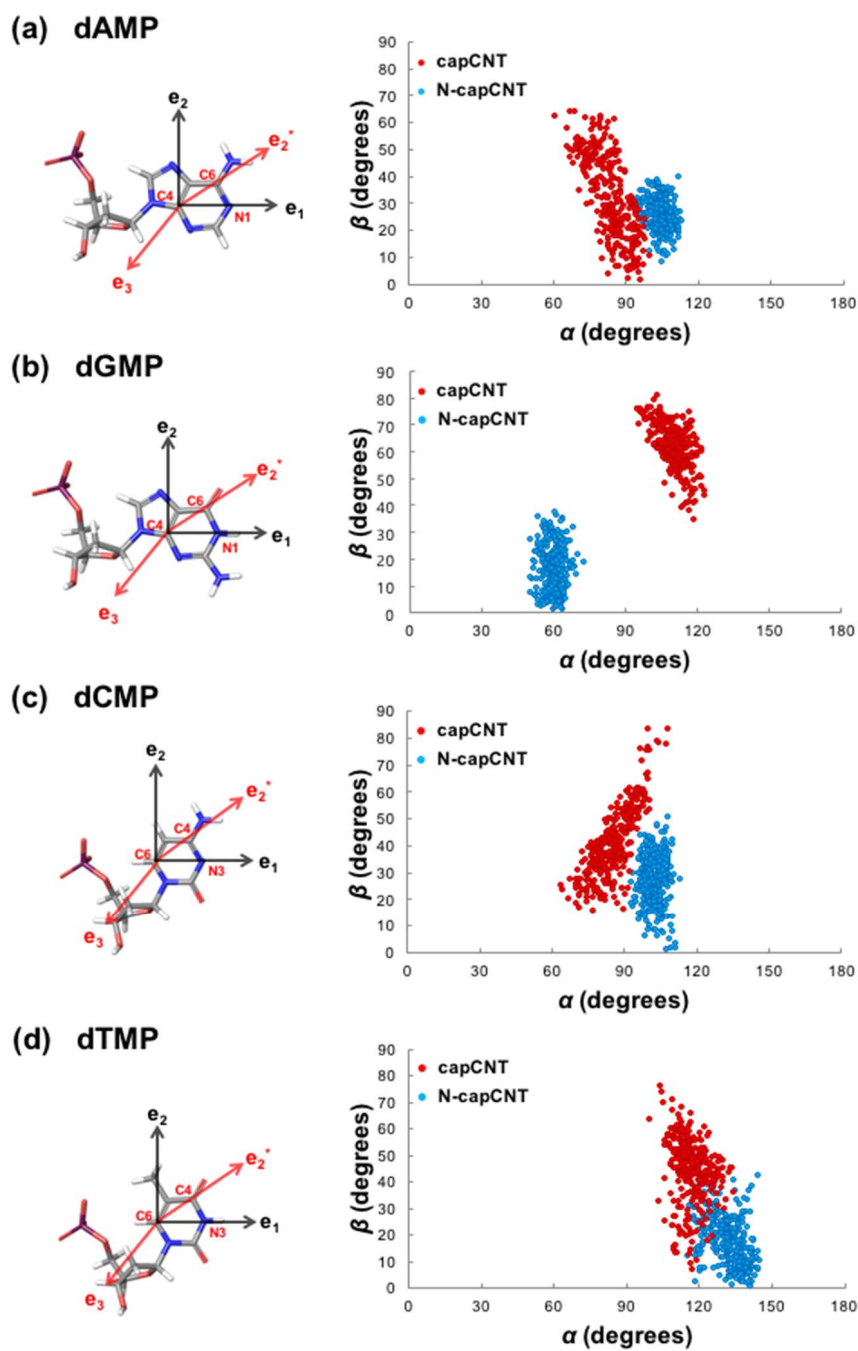

**Figure 3.** Definition of basis vectors used to specify the orientation of the four DNA bases (left panels), and corresponding scatter diagrams of angles α and β (right panels) in Set 1 simulations for the (a) dAMP, (b) dGMP, (c) dCMP, and (d) dTMP cases. On the right panels, red and blue dots indicate capCNT and N-capCNT cases, respectively.

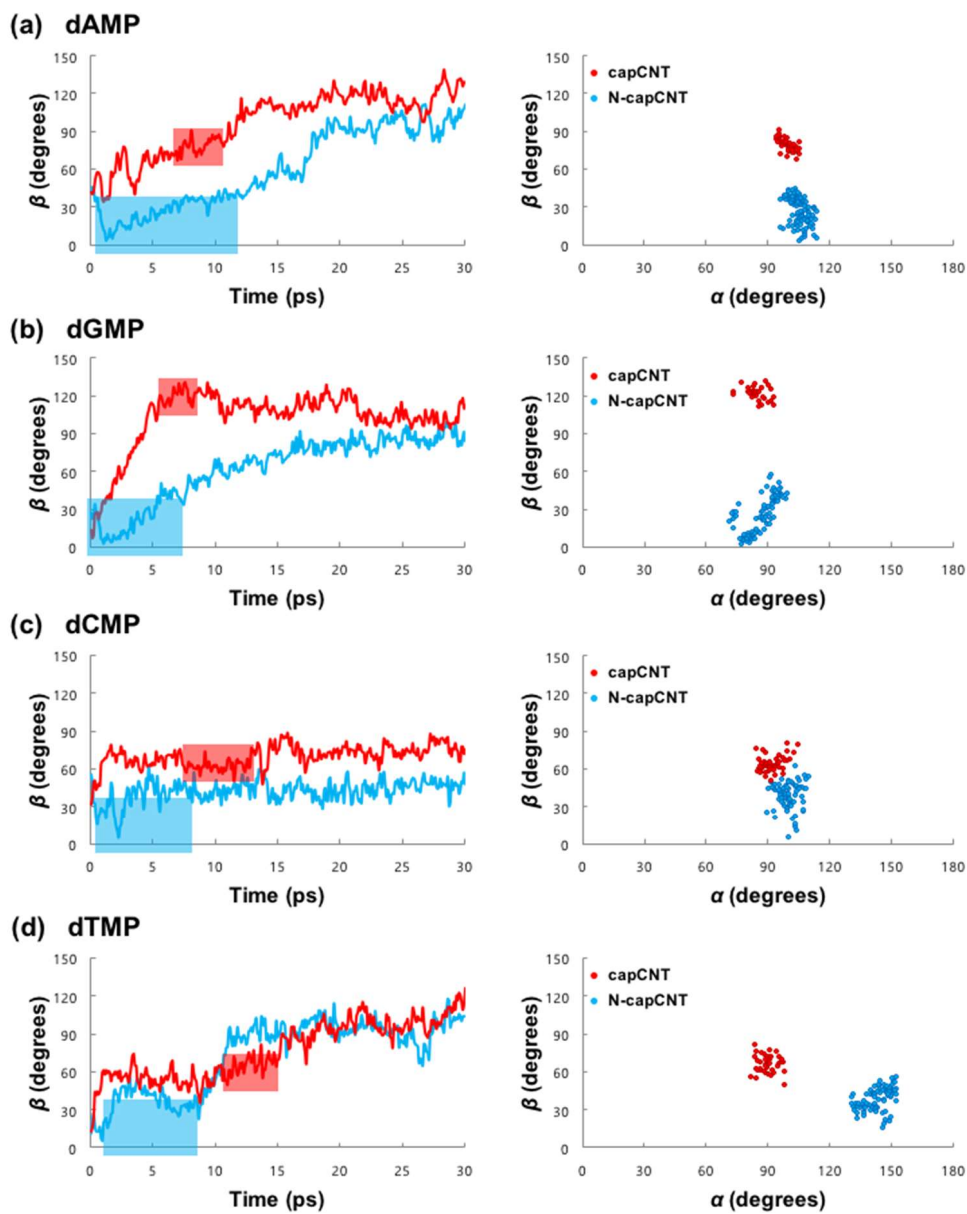

**Figure 4.** Time propagations of *β* for 30 ps (left panels) and scatter diagrams for tilt angles *α* and *β* (right panels) in *Set 2* simulations for the (a) dAMP, (b) dGMP, (c) dCMP, and (d) dTMP cases. On the left panels, the periods in which nucleobases are entrapped in the nanogap are indicated as red and blue boxes for the capCNTs and N-capCNTs, respectively. On the right panels, red and blue dots indicate the angles within the retention period for capCNTs and N-capCNTs, respectively.